\documentclass{ws-ijmpa}

\begin{document}

\markboth{Armando Bernui, Marcelo J. Rebou\c{c}as}
{Testing Gaussianity in the WMAP OT map}

%
\catchline{}{}{}{}{}
%

\title{TESTING GAUSSIANITY IN THE WMAP DATA OF OT FOREGROUND REDUCED MAP}

\author{ARMANDO BERNUI and MARCELO J. REBOU\c{C}AS}

\address{Centro Brasileiro de Pesquisas F\'{\i}sicas
\\ Rua Dr. Xavier Sigaud 150, \ 
22290-180 \ Rio de Janeiro -- RJ, Brazil \\
bernui@cbpf.br, reboucas@cbpf.br}

\maketitle

\begin{history}
\received{Day Month Year}
\revised{Day Month Year}
\end{history}

\begin{abstract}
A considerable effort has recently gone into the study of Gaussianity
of cosmic microwave background (CMB) data.
Among such attempts, there is one with two non-Gaussianity indicators, 
proposed by the authors, and used in a search for significant
deviation from Gaussianity in the WMAP internal linear combination 
(ILC) and in the single frequency WMAP maps with the $KQ75$ mask.  
Here we extend and complement these results by performing a similar analysis 
for the de Oliveira-Costa and Tegmark (OT) WMAP three-year $KQ75$ masked map, 
in which the foreground is reduced through a different statistical
cleaning procedure. 
\end{abstract}

\ccode{PACS numbers: 98.80.Es, 98.70.Vc, 98.80.-k}

\section{Introduction}  \label{Intro}

A key prediction of the so-called slow-roll inflationary models is that 
they cannot generate detectable non-Gaussianity of the cosmic microwave 
background (CMB) temperature fluctuations within the accuracy of the 
Wilkinson Microwave Anisotropy Probe (WMAP).\cite{Gauss_Single-field} 
There are, however, a number of inflationary models that predict 
non-Gaussianity  at a level detectable by the WMAP experiment (see 
the reviews Refs.~\refcite{Inflation-reviews}). 
Thus, non-Gaussianity in the CMB data may be potentially 
important to discriminate classes of inflationary models.

The WMAP team have found that the CMB data are consistent with 
Gaussianity~\cite{wmap1to5yr} 
However, a  number of recent analyses of CMB data performed 
with different statistical estimators have provided indications of 
deviations from Gaussianity in the CMB temperature fluctuations  
(for a list of references see, for example, Ref.\refcite{Bernui-Reboucas2008}
and references therein).  
On the other hand, one does not expect that a single statistical 
estimator can be sensitive to all possible forms of non-Gaussianity 
that may be present in WMAP data.  Thus, it is important to employ 
alternative indicators in order to  shed light on the possible causes 
for the reported non-Gaussianity of CMB data. 

Recently we proposed two new non-Gaussianity indicators, based on skewness 
and kurtosis  of large-angle patches of CMB maps, which provide a directional 
measure 
of deviation from Gaussianity on large angular scales.\cite{Bernui-Reboucas2008}
A distinctive feature of these indicators is that they make it possible to 
construct sky maps of Gaussianity from the CMB temperature data.
In Ref.~\refcite{Bernui-Reboucas2008} we also built and studied in details 
such maps of Gaussianity generated from the data of WMAP three and five-year 
internal linear combination (ILC) maps with $KQ75$ mask, and found no 
statistically  significant deviation from Gaussianity.%
\footnote{We have also used our indicators to test for Gaussianity the 
single frequency foreground unremoved WMAP three and five-year maps,
but here for  comparison we focus on the foreground-reduced ILC and OT maps.} 
Here we extend and complement these studies by using our indicators
to carry out a new analysis of Gaussianity with the CMB data of the 
so-called OT map\cite{OT} three-year $KQ75$ masked map, in which the 
Galactic foregrounds are cleaned with the TOH procedure.\cite{TOH}

\section{Indicators and Maps}  \label{Indicators}

The steps in the construction of our non-Gaussianity 
indicators and the associated maps, are the following:%
\footnote{For a detailed discussion of the indicator briefly 
presented here we refer the readers to Ref.~\refcite{Bernui-Reboucas2008}.}
\begin{romanlist}
\item[{\bf i.}]
Take a discrete set of points $\{j=1, \ldots ,N_{\rm c}\}$ homogeneously 
distributed on the CMB celestial sphere $S^2$ as the centers of spherical 
caps of a given aperture $\gamma$;
\item[{\bf ii.}]
Calculate for each cap $j$ the skewness  and kurtosis 
given, respectively,  by 
\begin{equation}
S_j   \equiv  \frac{1}{N_{\rm p} \,\sigma^3_{\!j} } \sum_{i=1}^{N_{\rm p}}
\left(\, T_i\, - \overline{T} \,\right)^3  
\quad \mbox{and} \quad 
K_j   \equiv  \frac{1}{N_{\rm p} \,\sigma^4_{\!j} } \sum_{i=1}^{N_{\rm p}}
\left(\,  T_i\, - \overline{T} \,\right)^4 - 3 \,,
\end{equation} 
where $N_{\rm p}$ is the number of pixels in the $j^{\,\rm{th}}$ cap,
$T_i$ is the temperature at the $i^{\,\rm{th}}$ pixel, $\overline{T}$ is
the CMB mean temperature, and $\sigma$ is the standard deviation.
Clearly, the numbers $S_j$ and $K_j$ obtained in this way for each cap
can be viewed as a measure of non-Gaussianity in the direction of
the center of the cap $(\theta_j, \phi_j)$.
\item[{\bf iii.}]
Patching together the $S_j$ and $K_j$ values for each cap, 
one obtains our indicators, i.e., discrete functions $S = S(\theta,\phi)$ 
and $K = K(\theta,\phi)$ defined over the celestial sphere, which can be 
used to measure the deviation from Gaussianity as a function of 
$(\theta,\phi)$.
\end{romanlist} 

Clearly, the discrete functions $S = S(\theta,\phi)$ and $K = K(\theta,\phi)$ 
can be expanded into their spherical harmonics in order to have their power 
spectra $S_{\ell}$ and $K_{\ell}$. 
Thus, for example, for the skewness one has
$S (\theta,\phi) = \sum_{\ell=0}^\infty \sum_{m=-\ell}^{\ell}
b_{\ell m} \,Y_{\ell m} (\theta,\phi)$ and 
$S_{\ell} = (2\ell+1)^{-1}\sum_m |b_{\ell m}|^2$. 
Similar expressions obviously  hold for the kurtosis $K = K(\theta,\phi)$.

\section{Main Results and Conclusions}

In this section we shall report the results of our calculation
of  $S = S(\theta,\phi)$ and $K = K(\theta,\phi)$ indicators
for the WMAP three-year data of the OT  with $KQ75$ mask, and
compare with the results of Ref.~\refcite{Bernui-Reboucas2008}
for the ILC WMAP three-year map with the same mask.

\begin{figure*}[htb!]
\begin{center}
\includegraphics[width=3.4cm,height=5.4cm,angle=90]{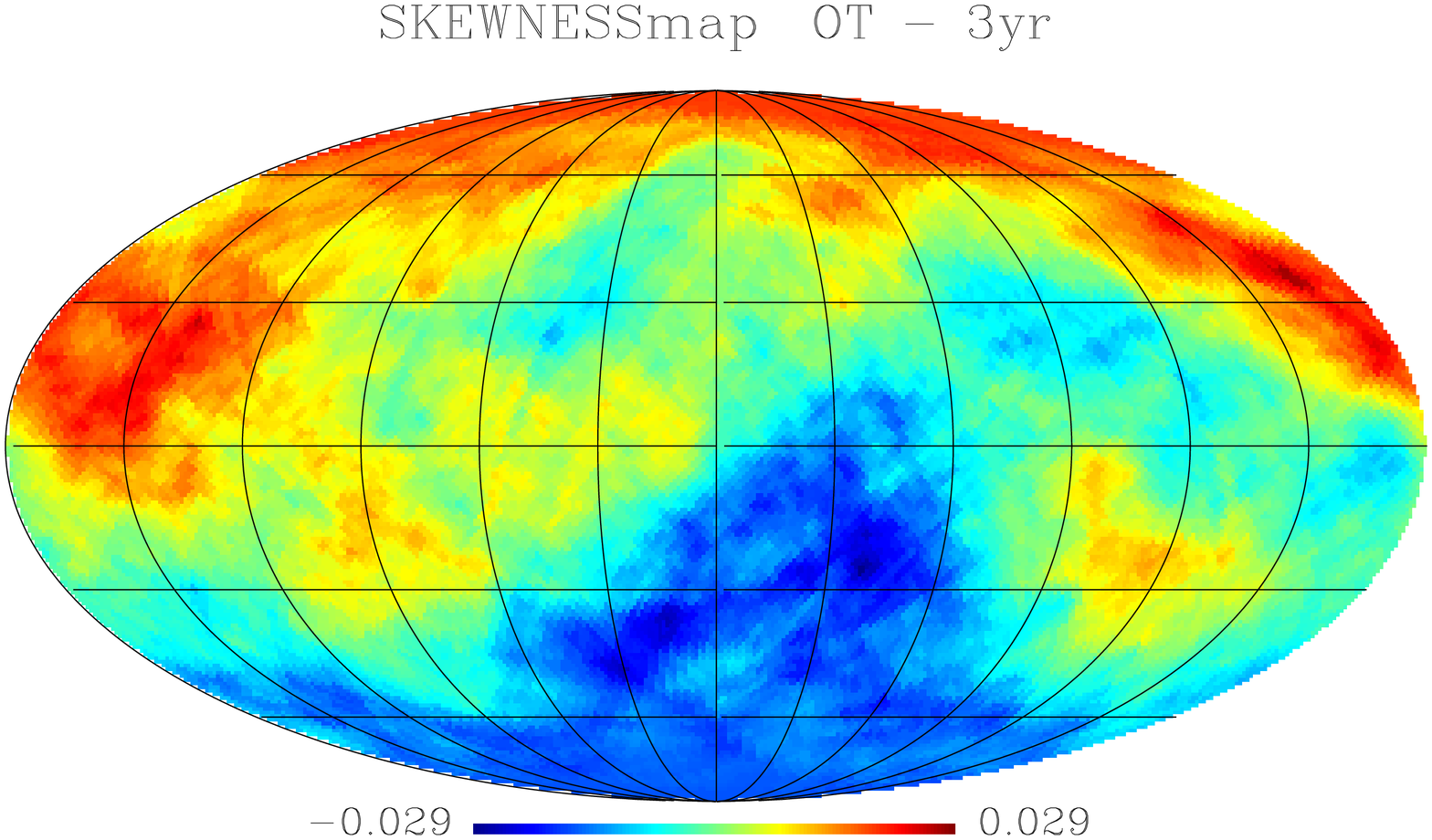}
\hspace{0.5cm}
\includegraphics[width=3.4cm,height=5.4cm,angle=90]{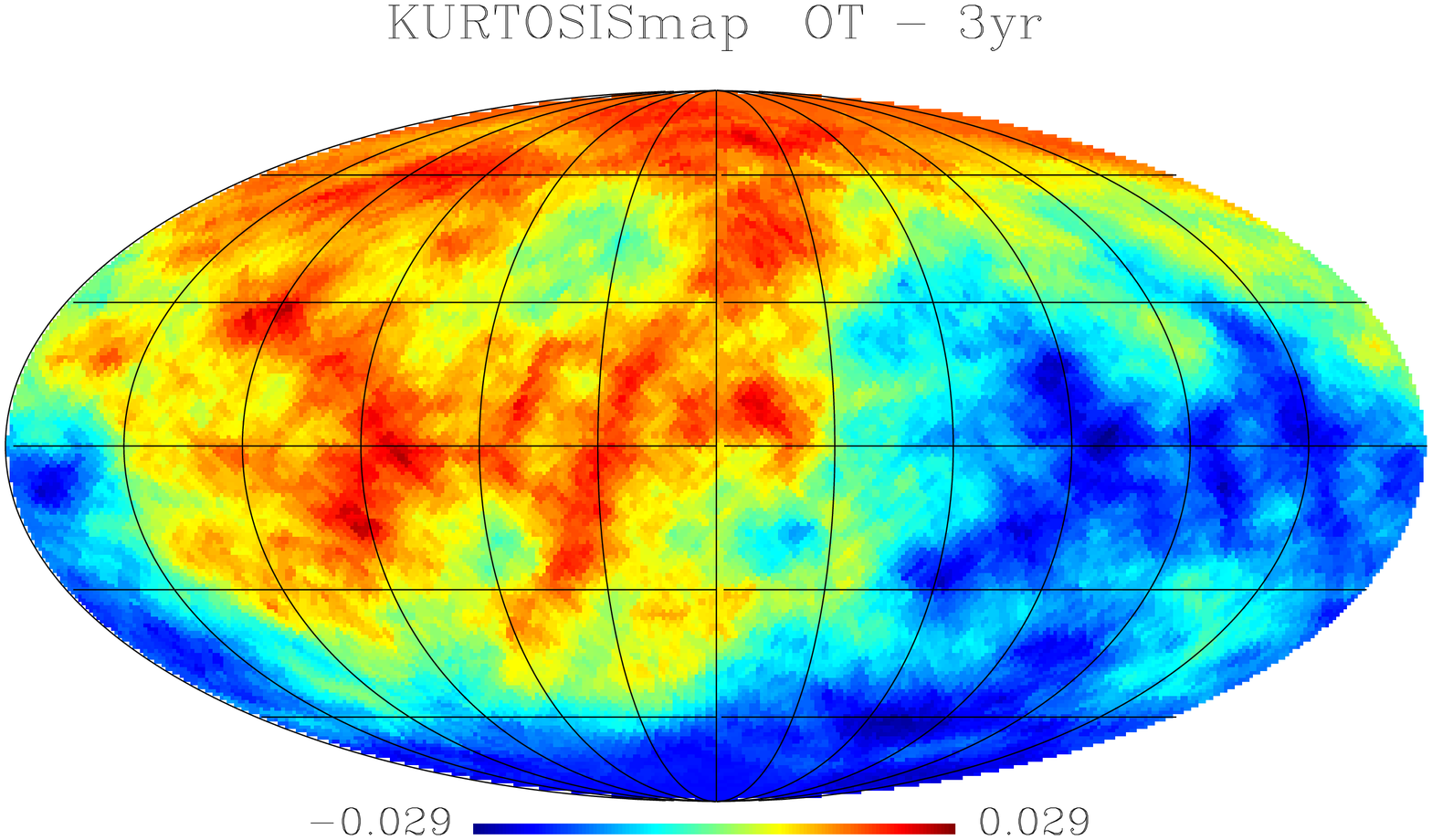}
\caption{\label{Fig1} Monopole removed Skewness (left) and monopole
removed Kurtosis (right)  maps generated from the WMAP three-year OT 
map with mask \emph{KQ75}. Colored high resolution versions of these 
figures are available in the arXiv version. \vspace{-0.6cm} } 
\end{center}
\end{figure*} 

In order to minimize the statistical noise, in the calculations 
of skewness and kurtosis indicator maps ($S-$map and $K-$map) 
we have scanned the celestial sphere with spherical caps of 
aperture  $\gamma = 90^{\circ}$, centered at $N_{\rm c}=12\,288$ 
points on the sphere homogeneously generated by using  HEALPix
package.\cite{Gorski-et-al-2005} 

Figure~\ref{Fig1} shows the Mollweide projection of the monopole
removed $S$ (left) and  $K$ (right) maps generated 
from the OT three-year map with a \emph{KQ75} mask. 
These maps show  spots with higher and lower values of indicators 
$S(\theta,\phi)$ and $K(\theta,\phi)$, thus suggesting at first sight 
non-Gaussianity of the CMB as given by OT temperature map.
Fig.~\ref{Fig1} is also suggestive of large-scale dominant components 
(low $\ell$ components) in the maps of both indicators.

\begin{figure*}[hbt!]
\begin{center}
\includegraphics[width=6.0cm,height=3.8cm]{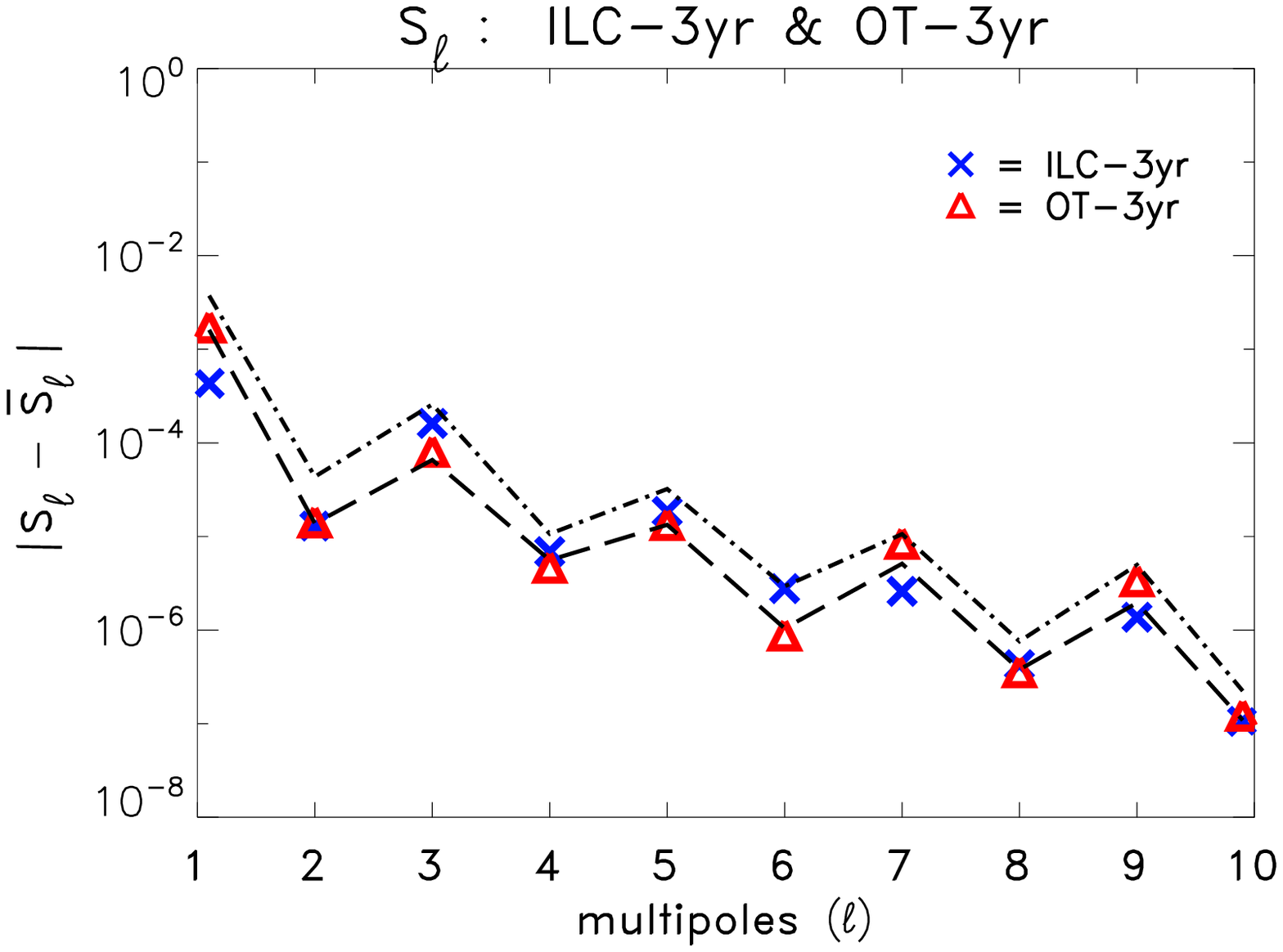}
\hspace{0.2cm}
\includegraphics[width=6.0cm,height=3.8cm]{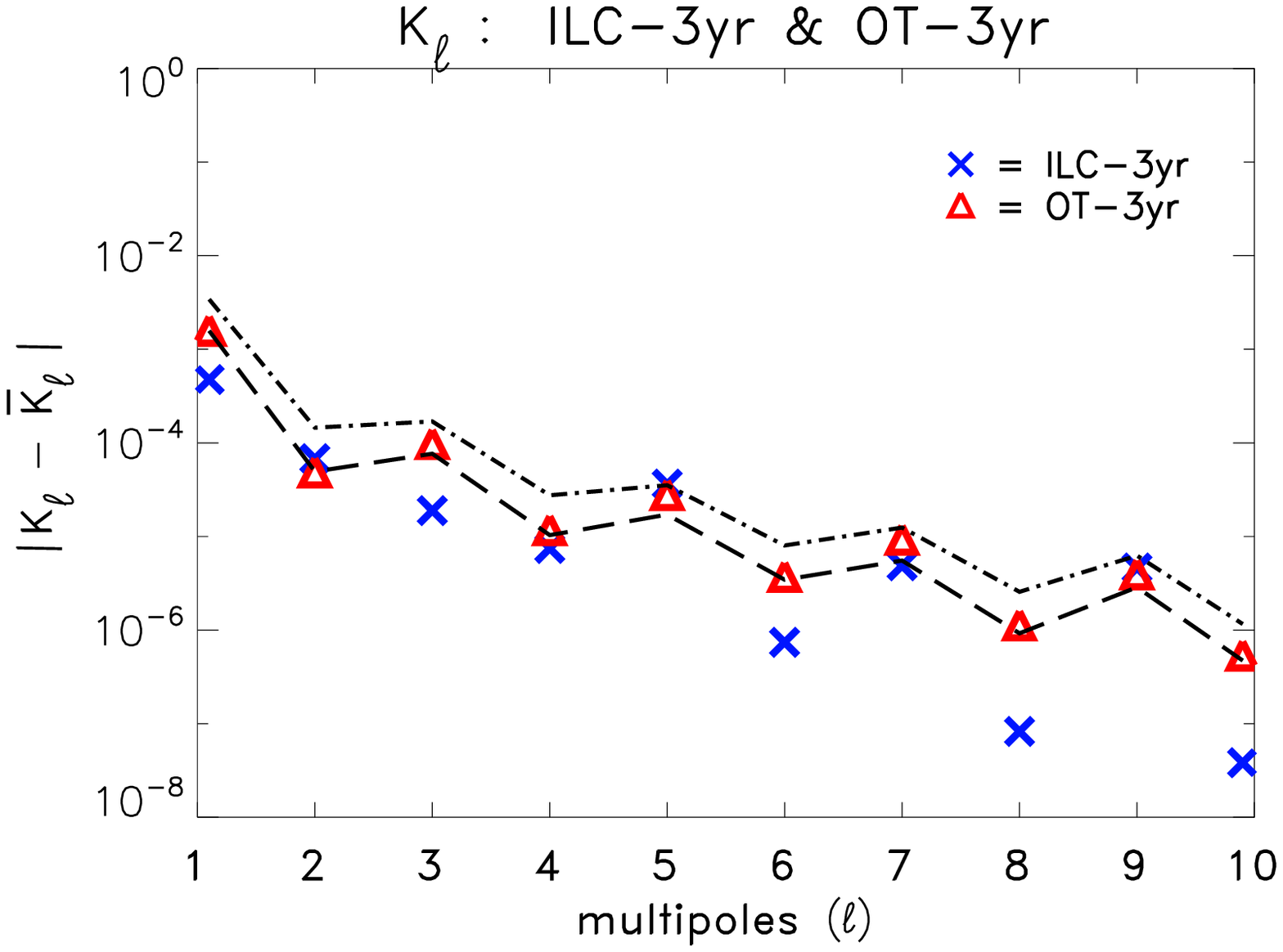}
\caption{\label{Fig2} Low $\ell $ \emph{differential} power spectra of skewness 
$|S_{\ell} - \overline{S}_{\ell}|$ (left) and kurtosis (right) 
$|K_{\ell} - \overline{K}_{\ell}|$
calculated from the monopole removed
maps of Fig.~\ref{Fig1}, which were generated from  the WMAP three-year OT 
map with mask \emph{KQ75}.  The $68\%$ and $95\%$ confidence
levels (obtained from MC simulated radom maps) are indicated, respectively, 
by the dashed and dash-dotted lines. \vspace{-0.6cm} }
\end{center}
\end{figure*}
%
In order to obtain quantitative information about the observed inhomogeneous
distribution for these non-Gaussianity indicators, we have calculated 
the power spectrum of both  
the $S-$map and $K-$map for the three-year OT \emph{KQ75} cut-sky map. 
These power spectra allow to estimate the statistical significance of the 
$S$ and $K$ multipole values, by comparing their power spectra with the 
averaged power spectrum of the $S$ and $K$ maps obtained by averaging over 
a set of $1\,000$ Monte-Carlo-generated statistically Gaussian CMB maps. In 
the left panel of Fig.~\ref{Fig2} we depict \emph{differential} power spectrum 
$|S_{\ell} - \overline{S}_{\ell}|$ (overline denotes the mean)
are obtained from the Monte-Carlo simulations
with each map being a stochastic realization of WMAP best-fitting angular 
power spectrum of the $\Lambda$CDM model, obtained by randomizing the multipole 
temperature components $a_{\ell m}$ within the cosmic variance limits. The right 
panel shows the corresponding plot $|K_{\ell} - \overline{K}_{\ell}|$
for the $K-$map. For a comparison the panels of Fig.~\ref{Fig2} also show  the
\emph{differential} power spectrum obtained from the WMAP three-year ILC 
\emph{KQ75} cut-sky map.

Figure~\ref{Fig2} shows the deviation of $S_{\ell}$ and 
$K_{\ell}$ values ($\ell=1,\!\cdots\!,\!10$) from the mean values of
$\overline{S}_{\ell}$ and $\overline{K}_{\ell}$ are  within (less than) 
$95\%$ the mean values (for each $\ell$) obtained from $1\,000$ 
Monte-Carlo-generated statistically Gaussian maps.
This clearly indicates how likely is the occurrence of the multipole values 
obtained from the OT data in the set of MC multipole values (from MC maps), 
giving therefore a clear indication of consistency with Gaussianity for the 
OT \emph{KQ75} masked map, in agreement with previous 
analyses performed with different statistical tools
(see, e.g. Refs.~\refcite{wmap1to5yr}--\refcite{Bernui-Reboucas2008}).

\section*{Acknowledgments}

This work is supported by Conselho Nacional de Desenvolvimento 
Cient\'{\i}fico e Tecnol\'{o}gico (CNPq) - Brasil, under grant No. 472436/2007-4.
M.J.R. and A.B. thank CNPq and PCI-INPE/CNPq for the grants under which this work
was carried out. We acknowledge use of the Legacy Archive for Microwave Background 
Data Analysis (LAMBDA) and the HEALPix package.\cite{Gorski-et-al-2005}



\begin{thebibliography}{00}    

\bibitem{Gauss_Single-field} 
V. Acquaviva, N. Bartolo, S. Matarrese and A. Riotto
\emph{Nucl. Phys. B} \textbf{667}, 119 (2003);
J. Maldacena, \emph{JHEP} \textbf{0305} 013 (2003), arXiv:astro-ph/0210603; 
M. Liguori, F. K. Hansen, E. Komatsu, S. Matarrese and A. Riotto,
\emph{Phys. Rev. D} \textbf{73}, 043505 (2006).

\bibitem{Inflation-reviews} B. A. Bassett, S. Tsujikawa and D. Wands, 
\emph{Rev. Mod. Phys.} \textbf{78}, 537 (2006);
A. Linde, \emph{Lect. Notes Phys.} \textbf{738}, 1 (2008).

\bibitem{wmap1to5yr}
E. Komatsu et al., \emph{Astrop. J. Suppl.} \textbf{148}, 119 (2003);
D. N. Spergel et al., \emph{Astrop. J. Suppl.} \textbf{170}, 377 (2007);
E. Komatsu et al., arXiv:0803.0547 [astro-ph]. See also
related refs.: B. Lew, \emph{JCAP} 08 (2008) 017;
B. Lew,  arXiv:0808.2867v1 [astro-ph] and references therein.

\bibitem{Bernui-Reboucas2008} A. Bernui and M. J. Rebou\c{c}as,
Searching for non-Gaussianity in the WMAP data, arXiv:0806.3758v1 [astro-ph]. 
See also the related references
A. Bernui, B. Mota, M. J. Rebou\c{c}as and R. Tavakol, 
\emph{Astron. Astrophys.} \textbf{464}, 479 (2007);
A. Bernui, B. Mota, M. J. Rebou\c{c}as and R. Tavakol, 
\emph{Int. J. Mod. Phys. D} \textbf{16}, 411 (2007). 

\bibitem{OT} A. de Oliveira-Costa and M. Tegmark, 
\emph{Phys. Rev. D} \textbf{74},  023005 (2006).  

\bibitem{TOH} M. Tegmark, A. de Oliveira-Costa and A. J. S. Hamilton, 
\emph{Phys. Rev. D} \textbf{68}, 123523  (2003). 


\bibitem{Gorski-et-al-2005} K. M. G\'orski, E. Hivon, A. J. Banday, 
B. D. Wandelt,  F. K. Hansen, M. Reinecke and M. Bartelman, 
\emph{Astrophys. J.} \textbf{622}, 759 (2005).

\end{thebibliography}
\end{document}